\documentclass[chicago]{emulateapj}

\shorttitle{TeV Blazar \emph{BVRI} Standards} \shortauthors{Pace et al.}

\usepackage{graphicx}
\usepackage{mathtools}
\usepackage{multirow}

\slugcomment{To appear in the PASP}

\shorttitle{Tev Blazar Standard Stars}
\shortauthors{Pace et al.}

\begin{document}


\title{The Stability of \emph{BVRI} Comparison Stars Near Selected TeV Blazars}

\author{Cameron J. Pace}

\affil{Department of Astronomy, Indiana University, Bloomington, IN 47404}
\email{cjpace@indiana.edu}

\author{Richard L. Pearson III}
\affil{Dept. of Physics \& Astronomy, University of Denver, Denver, CO, USA 80208}
\email{richard.pearson@du.edu, jmoody@physics.byu.edu}

\author{J. Ward Moody, Michael D. Joner, Bret Little}
\affil{Department of Physics and Astronomy, Brigham Young University, N484 ESC, Provo, Utah 84602}
\email{ jonerm@forty-two.byu.edu, bret.little@gmail.com}

\begin{abstract}

We have measured Johnson \emph{BV} and Cousins \emph{RI} magnitudes for comparison stars near five TeV blazars. We compare our values with published values, spanning 25 years in some cases, to identify those stars that are most likely proven stable. To avoid zero-point offsets mimicking long-term variability, we based our analysis on the standard deviation between measurements after a mean offset between data sets was removed. We found most stars to be stable at the 0.04 magnitude level.  We confirm two stars as variable and identify two others as possibly being variable. In each of the five fields there are at least two stars, and typically many more, that show no evidence of variability.
\end{abstract}

\keywords{AGN - comparison stars - optical photometry}

\section{Introduction}

The standard model of Active Galactic Nuclei (AGN) successfully explains blazars as high-energy jets directed toward us at a small angle to the line of sight. This orientation Doppler-enhances the emission intensity, contracts the variability time scales, and leads to apparent superluminal motion in ejected radio knots. In extreme cases we observe intensified synchrotron radiation from the radio to the UV–X-ray band and inverse-Compton radiation at higher energies, up to the TeV domain.

As pointed out by \cite{2011ApJ...736..131A} the details of the physical processes underlying blazar emission are still unknown. In their study of Mrk 421 both a hadronic model based on the Synchrotron-Proton Blazar model of \cite{Had1,Had2} and a leptonic model based on the 1-zone Synchrotron Self-Compton model described in \cite{Lep} were able to reproduce the observed spectral energy distribution, the less-constrained leptonic model agreeing marginally better with the data. A dissimilar distribution of matter in these models leads to different predictions regarding the speed with which emission intensity can vary. So a key to discriminating between these two scenarios is the rate of multi-wavelength variability. Optical variability studies therefore provide data that help constrain these competing models.

It is essential in optical variability studies to first establish well-calibrated secondary standards within the field of view (FOV). These standards should be tied back to known systems such as Johnson-Cousins \emph{UBVRI} to enable interobject comparisons. They must be known to be stable so that their own intrinsic variability is not ascribed to the blazar itself, especially in cases where there are only one or two appropriately bright standards in the field. Establishing these standards now enables current and future all-sky surveys such as the Large-Scale Synoptic Telescope \citep[LSST;][]{LSST} to confidently tie their measurements to historic data.

Well-measured standards exist for many of the brighter blazars, but the set is not complete. Finder charts with photometric standards for 92 prominent blazars compiled by the Whole Earth Blazar Telescope (WEBT) project are maintained at the University of Heidelberg\footnote{http://www.lsw.uni-heidelberg.de/projects/extragalactic/charts/}. In these data the epochs of most recent calibration span 20 years with observations often in only two or three filters. In the set as a whole, photometric indices exist for 19\% in \emph{U}, 57\% in \emph{B}, 60\% in \emph{V}, 78\% in \emph{R} and 55\% in \emph{I}.  Most missing indicies are for objects that are rarely, if ever, monitored. Of greater concern here is that few of the standards, including those of the commonly observed objects, have been verified since the mid to late 1990s. It is important for monitoring accuracy to periodically verify standard magnitudes, weed out slowly varying stars, and extend the calibrations to unmeasured filters as needed.

In their excellent study of AGN and blazar comparison standards, \cite{ESO2001} make the case for establishing many secondary standards per field since larger numbers reduce the statistical error. This is correct but only with regard to random errors; it does not improve the systematic error if all stars in the same field possess the same zero-point offset. Additionally, the overall error can increase if too many of the secondary stars are at large distances from the object of interest where flat-fielding errors tend to be greatest. On the other hand, if at least one star is proven to be unvarying and tied accurately back to a standard system, then the \emph{precision} of the photometry can be enhanced from comparing the object of interest to an ensemble of stars while the \emph{accuracy} can be determined by referencing the zero-point to the well-known star.

In this paper we examine the variability of standard stars surrounding five of the more significant TeV blazars: Mrk 421, H1426+428, Mrk 501, 1ES1959+650, and BL Lac. We describe our observations in Section 2.  In Section 3 we present our results and in Section 4 we discuss the conclusions.

\section{Observations}
The objects observed were taken from the northern blazars listed by \cite{2004NewAR..48..527H} as having been detected at TeV energies. Each field has between four and sixteen previously published calibrated $BVRI$ standards in a $15'$ square FOV.

Observations were made with the BYU Robotic Observatory for Variable Object Research (ROVOR) telescope \citep{ROVOR}. This telescope consists of an RC Optical 0.40 m Ritchey-Chr\'{e}tien optical tube on a Paramount ME German-equatorial mount. The CCD camera is an FLI ProLine PL003 with a 1024 x 1024 24 \micron~pixel SITe detector. The FOV is $23.4'$ on a side with a resolution of $1.37 ''$/pixel. The filter set was Johnson-Cousins \emph{BVRI} closely matched to the specifications defined by \cite{1976PASP...88..557B}.

Data were taken on thirteen photometric nights between June \& September 2009.  Primary standards from \cite{2009AJ....137.4186L} were observed up to 20 times each night and never less than  five times. Approximately one-third of the time at the telescope was spent observing the standards and a unique transformation was determined for each night.

\begin{figure}
   \figurenum{1}  \centerline{\includegraphics[width=5.0in]{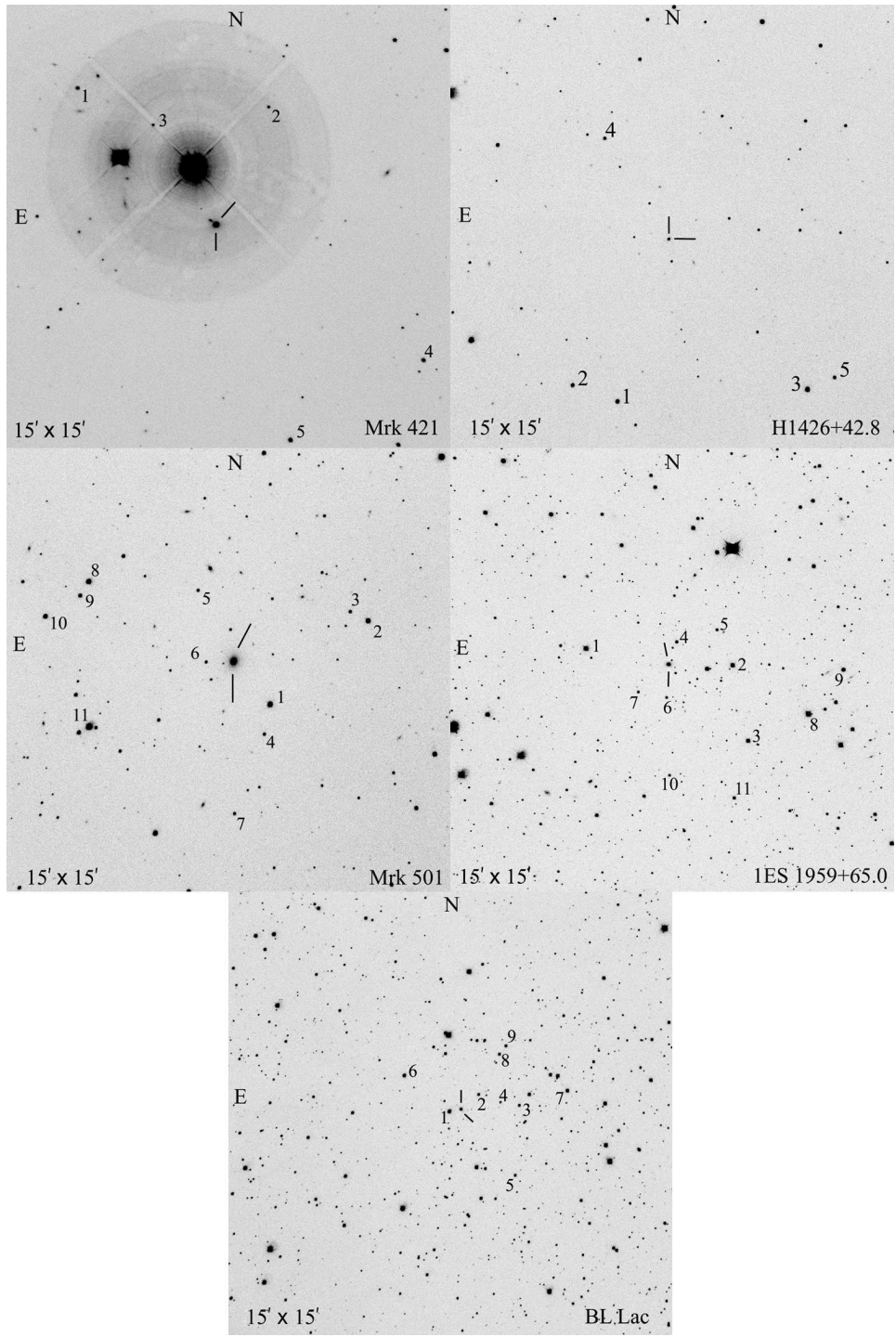}}
   \caption{\label{fig:f1} Finder charts for the blazars with comparison stars identified. North is to the top and east is to the left.  All charts are $15'$ x $15'$ as indicated.  All images were taken with the WMO $36''$ telescope in the V band.}
\end{figure}

All of our nightly observing routines followed closely the procedures outlined in \cite{2007ASPC..364...27L}. We began and ended each night by observing a Landolt field containing ten or so standard stars. We observed each standard field first near the meridian at an airmass no higher than 1.2 and every hour thereafter until reaching an airmass of 2.0.  At that time we began the process over with another standard field near the meridian. The blazar standards were always observed within the airmass range covered by the standards. We observed only in the western half of the sky with the telescope on the eastern side of the pier to avoid flat-fielding inconsistencies from the image inverting as the telescope crossed the pier. The CCD temperature was always at -30$^\circ$ C to eliminate errors in dark current calibrations caused by non-linearities with temperature. We observed in a ``palindrome'' sequence of \emph{VBRIIRBV} to place our mean observations in each filter at the same average air mass for a given setting. All of our images were 300 seconds in \emph{B}, 180 seconds \emph{V}, 90-120 seconds in \emph{R} and 30-60 seconds in \emph{I}.  Finder charts from images taken with the BYU West Mountain Observatory (WMO) $36''$ telescope are given in Fig. \ref{fig:f1}.

\subsection{Data Analysis}

We used two software programs to analyze our data: \texttt{iraf} and \texttt{Mira}. For about three-fourths of the nights \texttt{iraf} was used to apply the dark, bias, and flat-field frames as well as perform the aperture photometry. \texttt{Mira} was used for the remainder of the nights. To ensure that both programs produced similar results, two nights were reduced and photometered using both programs. The resulting instrumental magnitudes agreed to within 0.001 mag.  In either case we used an aperture radius of $5''$ and the stellar FWHM was typically on the order of $3''$.

Our observations were placed on a standard footing following the steps outlined by  \cite{2007ASPC..364...27L}. We discovered that the atmospheric extinction varied as a function of azimuth as well as altitude on many of our nights, an effect especially apparent in the \emph{I} band. This is because our observatory is near a desert and on a valley floor, making it more susceptible to errors from nightly extinction variation. To be cautious, we applied a unique atmospheric extinction correction to each program and standard field on each night. Since our program fields were always observed across an airmass of one to two, we could measure the extinction for each field through that range. We did this by first forming a magnitude versus airmass relation for each comparison star in a given field for each night.  We then found \emph{BVRI} extinction coefficients for each star in the field using a least-squares fit to this relation. The average of these coefficients was then used to remove the effects of extinction from all of the stars in that field. This process was repeated for all fields.

The extinction-corrected magnitudes were transformed onto the standard system using the equations
\begin{equation}
V\,=\,v_0 +\mu_{v}(B-V) +\zeta_{v},\nonumber
\end{equation}
\begin{equation}
B-V\,=\,\mu_{bv}(b-v)_0 +\zeta_{bv},\nonumber
\end{equation}
\begin{equation}
V-R\,=\,\mu_{vr}(v-r)_0 +\zeta_{vr},\nonumber
\end{equation}
\begin{equation}
V-I\,=\,\mu_{vi}(v-i)_0 +\zeta_{vi}.\nonumber
\end{equation}
The $\mu$ coefficients are the standard color terms, the $\zeta$ coefficients are the zero points, the capital letters are the literature values from \cite{2009AJ....137.4186L}, and the lower-case ``sub zero" terms are the instrumental magnitudes, corrected for extinction. Values for the coefficients were calculated using a least-squares fit to the extinction-corrected instrumental magnitudes and literature values of the Landolt standards.

\section{Results}

Magnitudes of the comparison stars, associated errors ($\sigma$), and number of nights the star was observed are presented in Table~\ref{tbl2}. In this table the names of blazars appear in the first column, while star numbers are in the second column. The next four columns give magnitudes and errors in the listed filters, while the final column lists the number of nights the star was observed. The errors quoted are the standard deviation of our final calibrated values.

\begin{figure}
   \figurenum{2}  \centerline{\includegraphics[width=6.0in]{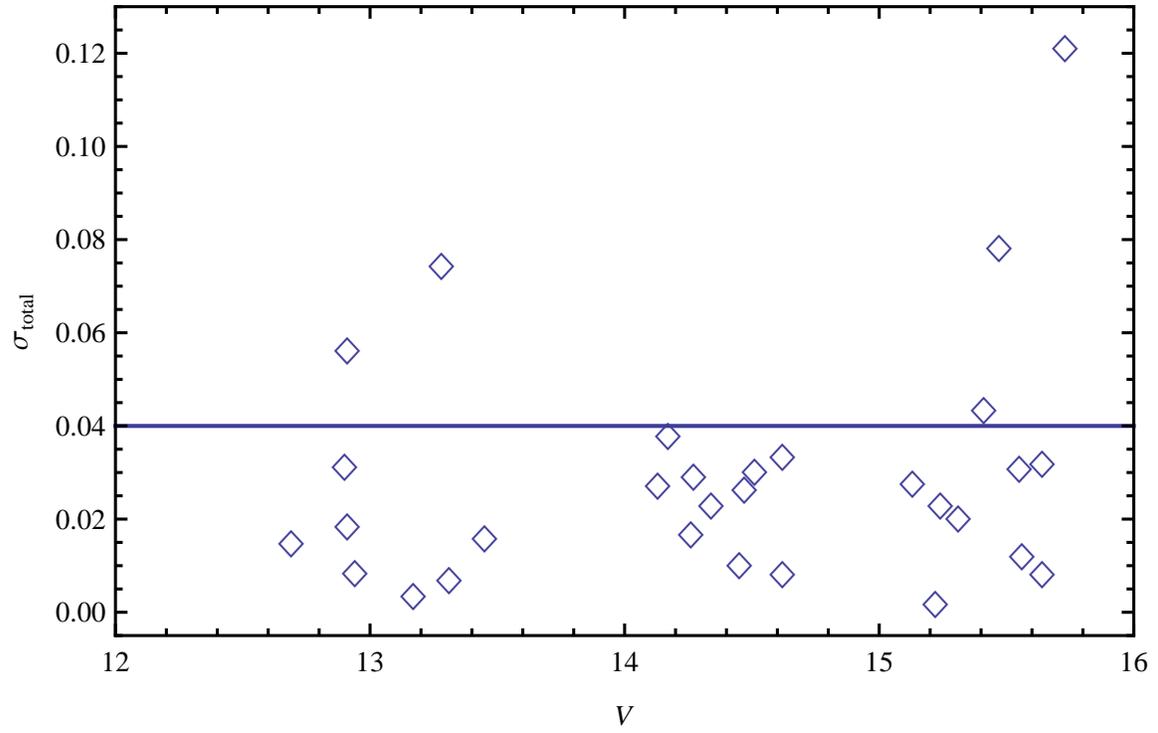}}
   \caption{\label{fig:f2} A plot of magnitude difference residuals against $V$ magnitude for all comparison stars considered in this paper. The larger the residual, the more likely the star is variable. The stars well above the nominal random error line of 0.04 magnitudes are Mrk 421 \#3 (0.122), H1426+428 \#4 (0.079), and 1ES1959+650 \#3 (0.075) and \#5 (0.056).}
\end{figure}

We estimated stellar variability by first finding the magnitude difference between our measurements and previously published measurements for each star. To minimize the effects of zero-point offsets, we found the average difference in each filter between our data and the data in a given publication and subtracted it off. We then averaged the absolute value of the residual of each star across all publications, which we plot against apparent $V$ magnitude in Fig. \ref{fig:f2}. These plotted points therefore are a normalized variation between publications.

The error of our measurements is on the order of 0.02 or 0.03 magnitudes which, again, includes the uncertainty of referencing back to the Landolt system. This is similar to the errors quoted in most of the previous studies against which we are comparing. Adding typical errors in quadrature gives a value of about 0.04 magnitudes which is a reasonable approximation of the expected value of an average point in Fig. \ref{fig:f2}. Stars with residuals above this value, shown as a horizontal line in that figure, are suspect for variability. One can see five stars that are suspect by this criterion. In the sections below we discuss these outliers individually.

The statistic formed above is better for finding variability than for determining stability since it has a mean offset removed as explained. To establish the stars of greatest stability, we felt it best to simply choose those with the closest agreement in published photometry across all references with no offset removed. However, we did not consider references with an offset larger than 0.04 magnitudes.

Stars of greatest estimated stability in each field are listed in Table~\ref{tbl1}. The names of blazars appear in the first column, while star numbers are in the second column. In columns three through six we give their $BVRI$ magnitudes and errors as determined by an error-weighted mean across all references. The errors thus calculated are from random sampling only, no attempt was made to formally estimate the systematic errors. We follow the numbering schemes of \cite{1998A&AS..130..305V} and \cite{1996A&AS..116..403F} in designating the comparison stars. Other identifications are listed in column 7 for clarity.

\subsection{MRK 421}

The star field around MRK 421 is sparse and dominated by the 6th magnitude star 51 UMa and its 12th magnitude companion BD$+$39 2414B. \cite{ESO2001} observed Mrk421 but there were no other stars inside their approximately $7'$ square FOV. There are, however, five published comparison stars outside this distance from \cite{1996A&AS..116..403F} \cite{2007Ap.....50...40D}.

Star 3 is problematic. It is the biggest outlier in Fig. \ref{fig:f2}. Our uncorrected \emph{BVR} values for it are 0.28, 0.04, and 0.15 magnitudes brighter than those presented by \cite{1998A&AS..130..305V}. This may be an intrinsic variability or just an effect from the orientation of our telescope secondary mirror spider arms, which, as can be seen in the finder chart, cause diffraction spikes from 51 UMa to cut through it. Either way, we recommend that star 3 not be used as a comparison star.

Excluding star 3, our values averaged across all filter bands agree to within 0.035 magnitudes with the average values of the other four stars presented by \cite{1996A&AS..116..403F} and within 0.006 magnitudes with \cite{2007Ap.....50...40D}. Since this agreement is within the published errors, we conclude that  stars 1,2, 4, and 5 show no evidence of variability.

\subsection{H1426+428}

The field near H1426+428 is also sparsely populated. Because of this, we have added only one new comparison star to the four presented by \cite{1991ApJS...77...67S}. 

With only one set of published values to compare against, this field cannot be examined in the same manner as the others, so we simply compare data. When averaging differences across all bands, the \cite{1991ApJS...77...67S} values differ from ours by -0.017 for \#1, 0.0625 for \#2, 0.038 for \#3, and 0.13 for \#4. The differences for stars \#1, and \#3, are within the errors or nearly so. Star \#3 shows a more significant difference but we cannot ascribe it to variability from this one data point. However, star \#4 is so different in value that it is hard to imagine how errors of observation by either \cite{1991ApJS...77...67S} or us could cause such a discrepancy. Even when removing the average difference between our two papers, the offset is still 0.07 magnitudes, the value we chose to plot in Fig. \ref{fig:f2}. Upon that discrepancy alone we recommend avoiding star 4.

\subsection{MRK 501}

Mrk 501 has a nicely populated field with eight published magnitudes from three sources \cite{1996A&AS..116..403F}, \cite{1998A&AS..130..305V}, and \cite{2005Ap.....48..304D}.  \cite{ESO2001} publish magnitudes for 18 faint stars encompassing the faintest ones near the object. We publish values for three more.

Stars 1 and 4 show excellent agreement across all authors. The difference between our values and the sources cited above are 0.033, 0.016, -0.004, and -0.004 for star 1 and -0.01, 0.01, 0.008, and -0.009 for star 4 respectively with the caveat that the B value for star 4 from \cite{1998A&AS..130..305V} was rejected as an obvious outlier. Being close to Mrk501, they are excellent references, especially the brighter star 1.

For star 2 the values from \cite{1996A&AS..116..403F} are internally consistent but off by 0.07 magnitudes from ours.  In \cite{2005Ap.....48..304D} they are again internally consistent but disagree with our values by 0.02 magnitudes. Star 3 shows the same trend; our values disagree with \cite{1996A&AS..116..403F} by 0.11 magnitudes and by 0.04 magnitudes with \cite{2005Ap.....48..304D}. These differences could be a zero-point offset or they might be intrinsic variation, particularly with star 3 since our agreement with \cite{2005Ap.....48..304D} data is usually much better than 0.04 magnitudes. Since the accuracy of these stars are in question, and stars 1 and 4 are so solid, we recommend referencing to stars 1 and 4.

\subsection{1ES1959+650}

1ES1959+650 has a well-populated field.  \cite{1998A&AS..130..305V} publish $B$ and $V$ values for seven stars and \cite{2007Ap.....50...40D} publish $BVRI$ values for the same set. We add four more stars.

Star 5 is actually the variable MM Draconis.  Its variability is apparent both in Fig. \ref{fig:f2} and in the large internal error of our data (see Table~\ref{tbl2}). We also confirm the finding of \cite{2007Ap.....50...40D} that star 3 is variable and should not be used. All remaining stars show no obvious variation.  The mean difference between our values and those of stars 1, 2, 4, 6, and 7 have remarkably good agreement with residuals of \cite{2007Ap.....50...40D}. These residuals averaged over all filters are 0.011, 0.007, 0.005, .001, and .013 respectively.  The good agreement is probably from the large number of observations in their study (33) and ours (10).

\subsection{BL Lac}

The field of BL Lac is also well populated and well calibrated, with a variety of stars close to the blazar. We consider here the comparison star magnitudes published by \cite{1985AJ.....90.1184S}, \cite{1996A&AS..116..403F}, and \cite{ESO2001}.

All standardized stars in this field show no obvious variation. We concur with \cite{ESO2001} that the photoelectric values from \cite{1985AJ.....90.1184S} are systematically brighter and exclude them from our analysis here and in Table ~\ref{tbl1}. The errors averaged across all filters are on the order of 0.02 magnitudes for all stars with star 5 having the largest variation of 0.035.

\section{Conclusion}

Most stars examined show no evidence of variability to within the errors of observation. Further, there are within each field at least two demonstrably stable standards stars, suitable for referencing. That these are often the brighter, closer stars unsurprisingly illustrates that minimizing error from photon statistics and flat-fielding produces the best data.

Referencing against many stars provides the most precise photometry. But for many applications a reference against the known stable stars will be sufficient and simpler. We note that Doroshenko et al. (2005, 2007) tied their photometry back to the Johnson-Cousins system through what they considered to be the most reliable measurements from other papers, particularly \cite{ESO2001}. We found that overall our numbers agreed best with theirs, often to within than 0.01 magnitudes.

\acknowledgments We are indebted to Alberto Sadun for many useful conversations and for acting as a mentor to several of us. We acknowledge the Brigham Young University, Department of Physics and Astronomy for their continued support of our research efforts. We are also grateful to the anonymous referee for several helpful suggestions. This work was supported by NSF grant AST-0618209.

\clearpage

\begin{deluxetable}{lllllll}
\tabletypesize{\scriptsize}
 \tablecolumns{7}
 \tablewidth{4.5in}
 \tablecaption{\emph{BVRI} magnitudes of comparison stars.  \label{tbl2}}
 \tablehead{
    \colhead{Blazar} & \colhead{Star} & \colhead{B($\sigma$)} &
    \colhead{V($\sigma$)} & \colhead{R($\sigma$)} & \colhead{I($\sigma$)}& \colhead{\emph{n}}}
 \startdata
&1&14.97 (0.02)&14.34 (0.02)&13.97 (0.03)& 13.63 (0.02)&2\\
&2&16.15 (0.03)&15.56 (0.05)&15.19 (0.05)&14.82 (0.05)&2\\
MRK 421&3&16.41 (0.05)&15.73 (0.06)&15.09 (0.06)&14.61 (0.04)&2\\
&4&15.17 (0.02)&14.13 (0.01)&13.52 (0.03)&13.05 (0.02)&2\\
&5&14.44 (0.03)&13.58 (0.01)&13.06 (0.03)&12.65 (0.01)&2\\
\tableline
&1& 15.60 (0.03)& 14.15 (0.02)& 13.26 (0.03)&  12.49   (0.02)&7\\
&2& 15.39 (0.04)& 14.55 (0.02)& 14.12 (0.03)& 13.73 (0.02)&7\\
H 1426+428&3& 14.19 (0.03)& 13.39 (0.02)& 12.97 (0.02)& 12.60 (0.02)&7\\
&4& 16.00 (0.07)& 15.39 (0.05)& 15.05 (0.04)& 14.72 (0.04)&7\\
&5& 15.98 (0.05)& 15.18 (0.03)& 14.74 (0.02)& 14.36 (0.03)&7\\
\tableline
&1& 13.54 (0.02)& 12.60 (0.02)& 12.08 (0.02)& 11.63 (0.02)&11\\
&2& 14.03 (0.02)& 13.17 (0.02)& 12.70 (0.02)& 12.28 (0.02)&11\\
&3& 15.87 (0.05)& 15.13 (0.03) & 14.69 (0.03)& 14.30 (0.03)&11\\
&4& 15.93 (0.04)& 15.31 (0.03)& 14.92 (0.02)& 14.57 (0.05)&11\\
&5& 16.14 (0.06)& 15.41 (0.05)& 14.97 (0.03)& 14.56 (0.03)&11\\
MRK 501&6& 16.77 (0.10)& 15.64 (0.04)& 14.91 (0.03)& 14.36 (0.04)&11\\
&7& 16.71 (0.07)& 15.44 (0.04)& 14.61 (0.02)& 13.94 (0.03)&11\\
&8& 13.48 (0.02)& 12.90 (0.02)& 12.55 (0.02)& 12.19 (0.02)&11\\
&9& 14.94 (0.03)& 14.34 (0.02)& 13.98 (0.02)& 13.63 (0.02)&11\\
&10& 14.37 (0.02)& 13.54 (0.02)& 13.03 (0.02)& 12.56 (0.02)&11\\
&11& 12.81 (0.02)& 11.82 (0.02)& 11.30 (0.02)& 10.83 (0.02)&11\\
\tableline
&1&13.34 (0.02)&12.69 (0.02)&12.28 (0.01)&11.90 (0.02)&10\\
&2&13.43 (0.03)&12.90 (0.01)&12.55 (0.01)&12.22 (0.02)&10\\
&3&15.02 (0.03)&13.30 (0.02)&12.33 (0.02)&11.47 (0.03)&10\\
&4&15.26 (0.03)&14.50 (0.01)&14.04 (0.02)&13.61 (0.03)&10\\
&5&15.65 (0.14)&14.64 (0.13)&14.00 (0.10)&13.43 (0.10)&10\\
1ES1959+650&6&15.94 (0.04)&15.21 (0.02)&14.77 (0.03)&14.38 (0.04)&10\\
&7&15.96 (0.04)&15.22 (0.02)&14.75 (0.03)&14.30 (0.03)&10\\
&8&12.69 (0.03)&12.21 (0.01)&11.88 (0.01)&11.57 (0.03)&10\\
&9&14.07 (0.02)&13.49 (0.01)&13.10 (0.01)&12.73 (0.03)&10\\
&10&16.19 (0.05)&15.27 (0.02)&14.72(0.03)&14.27 (0.03)&10\\
&11&15.14 (0.02)&14.38 (0.02)&13.93 (0.02)&13.53 (0.02)&10\\
\tableline
&1&14.65 (0.03)&12.94 (0.02)& 11.98 (0.03)& 11.15 (0.03)&10\\
&2&15.14 (0.03)&14.26 (0.03)& 13.77(0.02)& 13.31 (0.03)&10\\
&3&15.74 (0.04)&14.45 (0.03)& 13.72 (0.03)& 13.07 (0.03)&10\\
&4&16.33 (0.06)&15.54 (0.03)& 15.06 (0.03)& 14.58 (0.04)&10\\
BL Lac&5&15.48 (0.03)&14.46 (0.03)& 13.84 (0.02)& 13.30 (0.03)&10\\
&6&14.31 (0.03)&13.31 (0.03)& 12.71 (0.03)& 12.17 (0.03)&10\\
&7&13.99 (0.03)&13.28 (0.03)& 12.84 (0.03)& 12.41 (0.03)&10\\
&8&15.05 (0.03)&14.25 (0.03)& 13.77 (0.03)& 13.26 (0.03)&10\\
&9&15.07 (0.03)&14.16 (0.03)& 13.61 (0.03)& 13.07 (0.03)&10\\
\tableline
\enddata
\end{deluxetable}

\begin{deluxetable}{lllllll}
\tabletypesize{\scriptsize}
 \tablecolumns{7}
 \tablewidth{5.75in}
\tablecaption{Estimation of photometric values for demonstrably stable stars. \label{tbl1}}

\tablehead{
   \colhead{Field} & \colhead{Star} & \colhead{B($\sigma$)} &
   \colhead{V($\sigma$)} & \colhead{R($\sigma$)} & \colhead{I($\sigma$)}&
   \colhead{Other Designations} }
 \startdata
&1&14.982(0.009)&14.384(0.006)&14.022(0.004)&13.692(0.003)&\\
&2&16.173(0.010)&15.571(0.007)&15.200(0.004)&14.866(0.005)&3\tablenotemark{5} \\
MRK 421&4&15.135(0.010)&14.124(0.006)&13.534(0.004)&13.018(0.004)&8\tablenotemark{5} \\
&5&14.390(0.012)&13.571(0.006)&13.062(0.004)&12.628(0.004)&2\tablenotemark{5} \\
\tableline
&1&15.61(0.02)&14.16(0.01)&13.25(0.01)&12.46(0.01)&A\tablenotemark{2}\\
H 1426+428&2&15.45(0.02)&14.60(0.01)&14.15(0.01)&13.76(0.01)&B\tablenotemark{2}\\
&3&14.20(0.01)&13.40(0.01)&12.99(0.01)&12.61(0.02)
&C\tablenotemark{2}\\
\tableline
&1&13.540(0.003)&12.598(0.003)&12.083(0.003)&11.613(0.001)&5\tablenotemark{1}\\
MRK 501&4&15.961(0.018)&15.309(0.010)&14.931(0.009)&14.572(0.010)&3\tablenotemark{1}\\
\tableline
&1&13.359(0.012)&12.686(0.012)&12.280(0.004)&11.916(0.004)&\\
&2&13.446(0.014)&12.888(0.008)&12.534(0.004)&12.217(0.004)&\\
1ES1959+650&4&15.277(0.014)&14.501(0.008)&14.038(0.004)&13.619(0.004)&\\
&6&15.968(0.015)&15.204(0.012)&14.758(0.004)&14.365(0.004)&\\
&7&16.001(0.015)&15.225(0.012)&14.738(0.004)&14.315(0.004)&\\
\tableline
&1&14.643(0.020)&12.936(0.011)&11.966(0.014)&11.105(0.016)
&B\tablenotemark{3}\tablenotemark{4}, 12\tablenotemark{1}\tablenotemark{6}\\
&2&15.178(0.012)&14.278(0.008)&13.760(0.011)&13.313(0.013)
&C\tablenotemark{3}\tablenotemark{4}, 21\tablenotemark{1}\tablenotemark{6}\\
BL Lac&3&15.723(0.017)&14.452(0.004)&13.695(0.013)&13.024(0.009)
&H\tablenotemark{3}\tablenotemark{4}, 16\tablenotemark{1}\tablenotemark{6}\\
&4&16.380(0.021)&15.549(0.009)&15.051(0.009)&14.561(0.010)
&K\tablenotemark{3}\tablenotemark{4}, 17\tablenotemark{1}\tablenotemark{6}\\
&5&15.484(0.026)&14.439(0.007)&13.797(0.010)&13.237(0.013)
&3\tablenotemark{1}\tablenotemark{6}\\
&6&14.327(0.023)&13.298(0.004)&12.689(0.006)&12.161(0.015)
&25\tablenotemark{1}\tablenotemark{6}\\
&8&15.097(0.008)&14.254(0.016)&13.755(0.017)&13.262(0.020)
&28\tablenotemark{6}\\
&9&15.144(0.013)&14.191(0.020)&13.590(0.017)&13.070(0.023)
&29\tablenotemark{6}\\
\tableline
\enddata
 \tablecomments{(1) \cite{2005Ap.....48..304D}, (2)  \cite{1991ApJS...77...67S}, (3) \cite{1998A&AS..130..305V}, (4) \cite{1985AJ.....90.1184S}, (5) \cite{1969A&A.....3..436B}, (6) \cite{ESO2001}}

\end{deluxetable}

\end{document}